\documentclass[aps,pra,superscriptaddress,showpacs]{revtex4}
\usepackage{amsfonts}
\usepackage{amsmath}
\usepackage{amssymb}
\usepackage{graphicx}

\begin{document}

\title{Topological excitations in rotating Bose-Einstein condensates with
Rashba-Dresselhaus spin-orbit coupling in a two-dimensional optical lattice}
\author{Hui Yang}
\affiliation{Key Laboratory for Microstructural Material Physics of Hebei Province,
School of Science, Yanshan University, Qinhuangdao 066004, China}
\affiliation{Department of Physics, Xinzhou Teachers University, Xinzhou 034000, China}
\author{Qingbo Wang}
\affiliation{Key Laboratory for Microstructural Material Physics of Hebei Province,
School of Science, Yanshan University, Qinhuangdao 066004, China}
\affiliation{Department of Physics, Tangshan Normal University, Tangshan 063000, China}
\author{Ning Su}
\affiliation{Key Laboratory for Microstructural Material Physics of Hebei Province,
School of Science, Yanshan University, Qinhuangdao 066004, China}
\author{Linghua Wen}
\email{linghuawen@ysu.edu.cn}
\affiliation{Key Laboratory for Microstructural Material Physics of Hebei Province,
School of Science, Yanshan University, Qinhuangdao 066004, China}
\date{\today }

\begin{abstract}
We study the ground-state configurations and spin textures of rotating
two-component Bose-Einstein condensates (BECs) with Rashba-Dresselhaus
spin-orbit coupling (RD-SOC), which are confined in a two-dimensional (2D)
optical lattice plus a 2D harmonic trap. In the absence of rotation, a
relatively small isotropic 2D RD-SOC leads to the generation of ghost
vortices for initially miscible BECs, while it gives rise to the creation of
rectangular vortex-antivortex lattices for initially immiscible BECs. As the
strength of the 2D RD-SOC enhances, the visible vortices or the 2D
vortex-antivortex chains are created for the former case, whereas the
rectangular vortex-antivortex lattices are transformed into
vortex-antivortex rings for the later case. For the initially immiscible
BECs with fixed 2D RD-SOC strength, the increase of rotation frequency can
result in the structural phase transition from square vortex lattice to
irregular triangular vortex lattice and the system transition from initial
phase separation to phase mixing. In addition, we analyze the combined
effects of 1D RD-SOC and rotation on the vortex configurations of the ground
states for the case of initial phase separation. The increase of 1D SOC
strength, rotation frequency or both of them may result in the formation of
vortex chain and phase mixing. Furthermore, the typical spin textures for
both the cases of 2D RD-SOC and 1D RD-SOC are discussed. It is shown that
the system favors novel spin textures and skyrmion configurations including
an exotic skyrmion-half-skyrmion lattice (skyrmion-meron lattice), a
complicated meron lattice, a skyrmion chain, and a Bloch domain wall.
\end{abstract}

\maketitle

\section{Introduction}

The realization of Bose-Einstein condensates (BECs) is a milestone in the
study of ultracold atomic gases \cite{Dalfovo}. Owing to the unprecedented
level of control and precision, ultracold atomic gases provide an ideal test
ground to emulate various quantum phenomena in condensed matter physics \cite%
{Bloch,Zapf}. Recent experimental realization of artificial spin-orbit
coupling (SOC) \cite{Lin,Cheuk,ZWu,Huang,JRLi} which couples the internal
states and the orbit motion of the atoms not only offers a platform to
simulate the response of charged particles to external electromagnetic
field, but also give opportunities to search of novel quantum states \cite%
{Zhai,Ho,Sinha,YZhang,YLi,Kawakami,Stringari,Kartashov,Ruokokoski,XLi}.
Relevant investigations show that the SOC can lead to many new quantum
phases such as plane-wave phase \cite{CWang}, stripe phase \cite%
{XQXu,XZhou,HHu}, bright soliton \cite{Sakaguchi,Gautam}, dark soliton \cite%
{YXu}, half-quantum vortex configuration \cite{HHu,Ramach,XQLi}, and
topological superfluid phase \cite{ZWu}, which enrich the phase diagram and
physics of BEC system. In particular, the combined effects of SOC and
rotation\ on the BECs are predicted to generate various novel features.
Recently, Radi\'{c} \textit{et al.} \cite{Radic} has proposed an
experimental scheme for rotating spin-orbit-coupled BECs by using a suitable
control of the BECs. On the other hand, some groups have studied the
properties of BECs in various external potentials including harmonic trap
\cite{Aftalion,Fetter,Xu}, toroidal trap \cite{ACWhite,XFZhang},
concentrically coupled annular traps \cite{XFZhang2}, double-well potential
\cite{LWen,Javanainen,Kartashov2}, one-dimensional (1D) optical lattice \cite%
{JGWang}, and so on. It is demonstrated that the shape and dimension of the
external potential plays an important role in the stationary states and
dynamic properties of the BECs.

In this work, we investigate the topological excitations of rotating
two-component BECs with Rashba-Dresselhaus SOC (RD-SOC) in a two-dimensional
(2D) optical lattice plus a 2D harmonic trap. Actually, ultracold bosonic
gases loaded in a 2D optical lattice have attracted considerable interest.
By using two pairs of counterpropagating laser beams with orthogonal
polarization, a 2D optical lattice can be created \cite%
{Grynberg,Greiner,Clark}. Early investigations showed that the BECs in a
rotating optical lattice support interesting properties \cite{Pu,Tung}, such
as structural phase transition, domain formation, and vortex pinning, due to
the dynamically tunable periodicity and depth of the optical lattice. Recent
studies \cite{Radic2,DWZhang} demonstrated that SOC can significantly affect
the quantum phase transition of a spin-orbit-coupled bosonic gas in an
optical lattice from a superfluid to a Mott insulator, and may lead to some
novel magnetic phases, such as spiral phase and skyrmion crystals. Here we
show how the 2D isotropic RD-SOC, the 1D anisotropic RD-SOC and the rotation
frequency affect the ground-state structures and spin textures of BECs in a
2D optical lattice and a harmonic trap. In the absence of rotation, a small
2D RD-SOC can yield ghost vortices for the initially miscible BECs, while it
can induce the formation of vortex-antivortex lattices for the initially
immiscible BECs. For strong 2D RD-SOC, the visible vortices are generated in
the case of initial component mixing, while the vortex-antivortex rings are
formed in the case of initial component separation. When there exists
rotation driving, with the increasing rotation frequency a structural phase
transition from a square vortex lattice to a triangular vortex lattice
occurs for the initially immiscible BECs. In addition, the combined effect
of 1D RD-SOC and rotation on the ground state of the system for initially
immiscible BECs are discussed. It is found that the system supports novel
spin textures and topological defects including a peculiar
skyrmion-half-skyrmion lattice (skyrmion-meron lattice), a complicated meron
lattice, a skyrmion chain, and a Bloch domain wall.

This paper is organized as follows. In Sec. II, we present the theoretical
model of a rotating pseudospin-1/2 BEC with RD-SOC in a 2D optical lattice
and a harmonic trap. The topological structures and relevant spin textures
of the system are described and analyzed in Sec. III. Our findings are
summarized in Sec. IV.

\section{Model}

We consider a rotating quasi-2D pseudospin-$1/2$ BEC with RD-SOC in a 2D
optical lattice and a 2D harmonic trap. The Hamiltonian of the system can be
written as%
\begin{equation}
\hat{H}=\int dxdy\hat{\psi}^{\dagger }\left[ -\frac{\hbar ^{2}\triangledown
^{2}}{2m}+V(x,y)-\Omega L_{z}+v_{so}+\frac{g_{1}}{2}\hat{n}_{1}^{2}+\frac{%
g_{2}}{2}\hat{n}_{2}^{2}+g_{12}\hat{n}_{1}\hat{n}_{2}\right] \hat{\psi},
\label{Hamiltonian}
\end{equation}%
where $\hat{\psi}$ $=[\hat{\psi}_{1}(r),\hat{\psi}_{2}(r)]^{T}$ represents
collectively the spinor Bose field operators with 1 and 2 denoting spin-up
and spin-down, respectively. $\hat{n}_{1}=\hat{\psi}_{1}^{\dagger }\hat{\psi}%
_{1}$ and $\hat{n}_{2}=\hat{\psi}_{2}^{\dagger }\hat{\psi}_{2}$ are the
density operators of spin-up and spin-down atoms, respectively. Here we
assume that the two component atoms have the same mass $m$. The coefficients
$g_{j}$ $=\sqrt{8\pi }\hbar ^{2}a_{j}/ma_{z}$ $(j=1,2)$ and $g_{12}$ $=\sqrt{%
8\pi }\hbar ^{2}a_{12}/ma_{z}$ denote the intra- and interspecies
interaction strengths characterized by the $s$-wave scattering lengths $%
a_{j} $ and $a_{12}$ between intra- and intercomponent atoms, and $a_{z}=%
\sqrt{\hbar /m\omega _{z}}$ is the oscillation length in the $z$ direction.
For simplicity, we assume $g_{1}$ $=g_{2}$ $=$ $g$ throughout this paper.
The RD-SOC is given by $v_{so}=-i\hbar (k_{x}\hat{\sigma}_{x}\partial
_{x}+k_{y}\hat{\sigma}_{y}\partial _{y})$ \cite{Goldman}, where $\hat{\sigma}%
_{x}$ and $\hat{\sigma}_{y}$ are Pauli matrices, and $k_{x}$ and $k_{y}$
denote the SOC strength in the $x$ and $y$ directions. $\Omega $ is the
rotation frequency along the $z$ direction, and $L_{z}=-i\hbar (x\partial
_{y}-y\partial _{x})$ is the $z$ component of the angular-momentum operator.
The combined potential of a 2D optical lattice and a 2D harmonic trap is
expressed as \cite{Pu,Kartashov3}%
\begin{equation}
V\left( x,y\right) =V_{0}\left[ \sin ^{2}\left( \frac{2\pi x}{\lambda }%
\right) +\sin ^{2}\left( \frac{2\pi y}{\lambda }\right) \right] +\frac{1}{2}%
m\omega _{\perp }^{2}\left( x^{2}+y^{2}\right) ,  \label{Potential}
\end{equation}%
where $V_{0}$ denotes the depth of the optical lattice, which can be
controlled by the intensity of retroreflected laser beams, $\lambda $ is the
wavelength of the retroreflected laser beams, and $\omega _{\perp }$ is the
radial trap frequency. Based on mean-field theory, the Gross-Pitaevskii (GP)
energy functional of the system is given by
\begin{eqnarray}
E &=&\int dxdy[{\psi _{1}^{\ast }\left( -\frac{\hbar ^{2}}{2m}\nabla
^{2}+V\right) \psi _{1}+\psi _{2}^{\ast }\left( -\frac{\hbar ^{2}}{2m}\nabla
^{2}+V\right) \psi _{2}-\Omega \psi _{1}^{\ast }L_{z}\psi _{1}-\Omega \psi
_{2}^{\ast }L_{z}\psi _{2}}  \notag \\
{\newline
} &&{+\psi _{1}^{\ast }\hbar \left( -ik_{x}\partial _{x}-k_{y}\partial
_{y}\right) \psi _{2}+\psi _{2}^{\ast }\hbar \left( -ik_{x}\partial
_{x}+k_{y}\partial _{y}\right) \psi _{1}+\frac{g}{2}\left( \left\vert \psi
_{1}\right\vert ^{4}+\left\vert \psi _{2}\right\vert ^{4}\right)
+g_{12}\left\vert \psi _{1}\right\vert ^{2}\left\vert \psi _{2}\right\vert
^{2}].}  \label{EnergyFunctional}
\end{eqnarray}%
In our calculation, it is convenient to make the following parameter
transformations $\tilde{x}=x/a_{0}$, $\tilde{y}=y/a_{0}$, $\tilde{t}=\omega
_{\perp }t$, $\tilde{V}\left( x,y\right) =V\left( x,y\right) /\hbar \omega
_{\perp }$, $\tilde{\Omega}=\Omega /\omega _{\perp }$, $\beta =gN/\hbar
\omega _{\perp }a_{0}^{2}$, $\beta _{12}=g_{12}N/\hbar \omega _{\perp
}a_{0}^{2}$, $\tilde{k}_{q}=k_{q}/\omega _{\perp }a_{0}$ $(q=x,y)$, and $%
\tilde{\psi}_{j}=\psi _{j}a_{0}/\sqrt{N}$ $(j=1,2)$, where $a_{0}=\sqrt{%
\hbar /m\omega _{\perp }}$ is the characteristic length of the harmonic
trap, and $N=\int (\left\vert \psi _{1}\right\vert ^{2}+\left\vert \psi
_{2}\right\vert ^{2})dxdy$ is the number of atoms. In terms of the
variational method, we obtain the dimensionless coupled 2D GP equations,%
\begin{eqnarray}
i\partial _{t}\psi _{1} &=&\left[ -\frac{1}{2}\nabla ^{2}+V+\beta \left\vert
\psi _{1}\right\vert ^{2}+\beta _{12}\left\vert \psi _{2}\right\vert
^{2}-i\Omega \left( y\partial _{x}-x\partial _{y}\right) \right] \psi _{1}
\notag \\
&&+\left( -ik_{x}\partial _{x}-k_{y}\partial _{y}\right) \psi _{2},
\label{GPE1} \\
i\partial _{t}\psi _{2} &=&\left[ -\frac{1}{2}\nabla ^{2}+V+\beta \left\vert
\psi _{2}\right\vert ^{2}+\beta _{12}\left\vert \psi _{1}\right\vert
^{2}-i\Omega \left( y\partial _{x}-x\partial _{y}\right) \right] \psi _{2}
\notag \\
&&+\left( -ik_{x}\partial _{x}+k_{y}\partial _{y}\right) \psi _{1},
\label{GPE2}
\end{eqnarray}%
where the tilde is omitted for simplicity, and the dimensionless external
potential with $a=2\pi a_{0}/\lambda $ reads%
\begin{equation}
V=V_{0}\left[ \sin ^{2}\left( ax\right) +\sin ^{2}\left( ay\right) \right] +%
\frac{1}{2}\left( x^{2}+y^{2}\right) .  \label{DimensionlessPotential}
\end{equation}

In order to better understand the physical properties of this system, we use
the nonlinear Sigma model \cite{Mizushima,Kasamatsu,Kasamatsu1} and
introduce a normalized complex-valued spinor $\mathbf{\chi }=\left[ \chi
_{1},\chi _{2}\right] ^{T}$ with $\left\vert \chi _{1}\right\vert
^{2}+\left\vert \chi _{2}\right\vert ^{2}=1$. The main idea of the nonlinear
Sigma model is that pseudospin representation of the order parameter of a
system with internal degrees of freedom is userful to obtain a physical
understanding by mapping the system to a magnetic system. In this context,
two-component BECs can be treated as a spin-$1/2$ BEC. An exact mathematical
correspondence can be established between the two systems, where $\psi _{1}$(%
$\psi _{2}$) corresponds to the up (down) component of the spin-$1/2$
spinor. The detailed discussion can be referred to Refs. \cite%
{Mizushima,Kasamatsu,Kasamatsu1}. The two-component wave functions can be
expressed as $\psi _{1}=\sqrt{\rho }\chi _{1}$ and $\psi _{2}=\sqrt{\rho }%
\chi _{2}$, where $\rho =\left\vert \psi _{1}\right\vert ^{2}+\left\vert
\psi _{2}\right\vert ^{2}$ is the total density of the system. In the
pseudospin representation, the spin density is given by $\mathbf{S=\bar{\chi}%
\sigma \chi \ }$in which $\mathbf{\sigma }=(\sigma _{x},\sigma _{y},\sigma
_{z})$ are the pauli matrices. The components of $\mathbf{S}$ can be written
as \cite{Aftalion,WHan2,CFLiu}%
\begin{eqnarray}
S_{x} &=&2\left\vert \chi _{1}\right\vert \left\vert \chi _{2}\right\vert
\cos \left( \theta _{1}-\theta _{2}\right) ,  \label{SpindensityX} \\
S_{y} &=&-2\left\vert \chi _{1}\right\vert \left\vert \chi _{2}\right\vert
\sin \left( \theta _{1}-\theta _{2}\right) ,  \label{SpindensityY} \\
S_{z} &=&\left\vert \chi _{1}\right\vert ^{2}-\left\vert \chi
_{2}\right\vert ^{2},  \label{SpindensityZ}
\end{eqnarray}%
where $\theta _{j}$ $(j=1,2)$ is the phase of component wave function $\psi
_{j}$.

\section{Ground-state structures and spin textures for the case of 2D RD-SOC}

In our calculations, we numerically solve the GP Eqs. (\ref{GPE1}) and (\ref%
{GPE2}) and obtain the ground-state structure of the system by using the
imaginary-time propagation algorithm based on the Peaceman-Rachford method
\cite{Peaceman,LWen2}. We consider the isotropic 2D RD-SOC and 1D RD-SOC
effects on the ground states of rotating BECs in an optical lattice plus a
harmonic potential. In the present work, the parameters of the optical
lattice are chosen as $V_{0}=70$ and $a=4$, and the intra- and interspecies
interactions are both assumed to be repulsive. For the convenience of
discussion, when $\beta ^{2}<\beta _{12}^{2}$ $(\beta ^{2}>\beta _{12}^{2})$
we call the regime briefly initial component separation (initial component
mixing).

\begin{figure}[tbp]
\centerline{\includegraphics*[width=7.2cm]{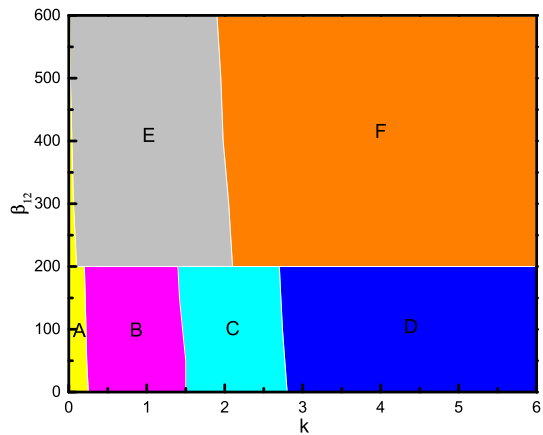}}
\caption{(Color online) Ground-state phase diagram of nonrotating
spin-orbit-coupled spin-$1/2$ BEC in an optical lattice plus a harmonic trap
with respect to $k$ ($k_{x}=k_{y}=k$) and $\protect\beta _{12}$ for $\protect%
\beta =200$. There are six different phases marked by A-F.}
\label{Figure1}
\end{figure}

Firstly, we discuss the effect of isotropic 2D RD-SOC on the ground-state
structure of spin-$1/2$ BEC without rotation in an optical lattice and a
harmonic trap. Relevant studies \cite{Zhai,Ho,Sinha,YZhang,YLi} showed that
the mean-field ground state for a nonrotating spin-orbit-coupled spin-$1/2$
BEC in a harmonic trap has two typical phases, plane-wave phase (i.e.
Thomas-Fermi (TF) phase) and stripe phase, depending on the nonlinear
interactions. In Fig. 1, we give the ground-state phase diagram of
nonrotating spin-orbit-coupled spin-1/2 BEC loaded in an optical lattice
plus a harmonic trap with respect to the isotropic SOC strength $k$ ($%
k_{x}=k_{y}=k$) and the interspecies interaction $\beta _{12}$. There are
six different phases marked by A-F, which differ in terms of their density
and phase distributions. In the following discussion, we will give a
detailed description of each phase. The density and phase profiles of the
B-F in Fig.1 are shown in Figs. 2(a)-2(e), respectively, where the
interaction parameters are $\beta =$ $200$, $\beta _{12}$ $=50$ for the
first three columns, and $\beta _{12}$ $=300$\ for the last two columns. The
isotropic 2D RD-SOC strengths are $k_{x}=k_{y}=0.5$ (a, d), $k_{x}=k_{y}=2$
(b), $k_{x}=k_{y}=5$ (c), and $k_{x}=k_{y}=2.5$ (e).\ Note that the odd and
even rows present component 1 and component 2, respectively. We start from
the case where SOC is sufficiently weak, which is indicated by the region A
in Fig. 1. The A phase is a periodically modulated TF phase in which no
vortex exists due to the very small isotropic RD-SOC (the density and phase
profiles are not shown here for the sake of simplicity). In the case of
relatively weak SOC, the system supports the B phase for $\beta _{12}<200$
and the E phase for $\beta _{12}>200$, whose ground states are shown in
Figs. 2(a) and 2(d), respectively. The ground state of the system exhibits
obvious phase mixing in Fig. 2(a), where several ghost vortices are
generated in the outskirts of each component (see the bottom two rows in
Fig. 2(a)) and they carry no angular momentum \cite{LWen,Kasamatsu2,LHWen}.
It is known that there are three fundamental types of vortices in cold atom
physics: visible vortex, ghost vortex, and hidden vortex. The visible vortex
is the ordinary quantized vortex which is visible in both the density
distribution and the phase distribution and carries angular momentum \cite%
{Fetter2}. For the ghost vortex, it shows up in the phase distribution as a
phase singularity and has no visible vortex core in the density distribution
and carries no angular momentum \cite{Kasamatsu2}. Ghost vortices can be
detected by the interference between two BECs, at least one of which
contains ghost vortices. Like ghost vortex, the hidden vortex as a phase
defect is invisible in the density profile but it carries angular momentum.
Only after including the hidden vortices can the well-known Feynman rule be
satisfied \cite{LWen,LHWen}, and the hidden vortices can be observed by
using the scheme of free expansion. In Fig. 2(d) the two component densities
display evident phase separation, where the topological defects in
individual components composed of alternately arranged vortices (clockwise
rotation) and antivortices (anticlockwise rotation) form rectangular
vortex-antivortex lattices.

\begin{figure}[tbp]
\centerline{\includegraphics*[width=12cm]{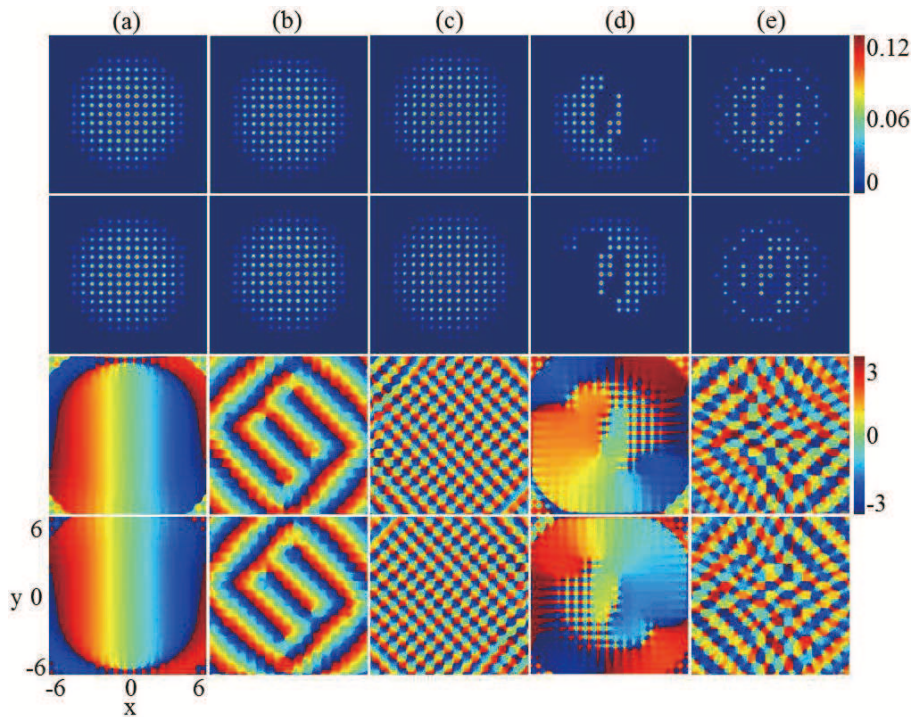}}
\caption{(Color online) Ground-state density distributions (the top two
rows) and phase distributions (the bottom two rows) for nonrotating spin-1/2
BEC with isotropic 2D RD-SOC in an optical lattice plus a harmonic
potential. (a) $k_{x}=k_{y}=0.5$, (b) $k_{x}=k_{y}=2$, (c) $k_{x}=k_{y}=5$,
(d) $k_{x}=k_{y}=0.5$, and (e) $k_{x}=k_{y}=2.5$. The interaction parameters
are $\protect\beta =200$, $\protect\beta _{12}=50$ for (a)-(c), and $\protect%
\beta _{12}=300$ for (d)-(e). The odd and even rows correspond to component
1 and component 2, respectively. The unit length is $a_{0}$.}
\label{Figure2}
\end{figure}

With the increase of isotropic RD-SOC strength, for the case of initial
component mixing the C phase emerges as the ground state of the system, as
shown by the region C in Fig. 1. The typical density and phase distributions
of this phase are given in Fig. 2(b), where the ground-state density
distributions of the system are similar to those in Fig. 2(a), but the ghost
vortices disappear and ordinary visible vortices occur. The main reason is
that the increased SOC offers more energy and angular momentum to the
system. As the SOC is further increased, the D phase emerges, as displayed
by the region D in Fig. 1. The density and phase distributions are shown in
Fig. 2(c). The phase consists of 2D vortex-antivortex chain where vortices
and antivortices form alternately arranged 2D chain structure in space. For
the case of initial phase separation, with the increase of isotropic RD-SOC
strength, the E phase transforms to the F phase in Fig. 1. Typical density
and phase distributions of such a phase are shown in Fig. 2(e). The
component densities keep separated (see the upper two rows in Fig. 2(e)) but
the topological defects evolve into vortex-antivortex rings in which the
vortices and antivortices are arranged alternately to form ring structures
(see the lower two rows in Fig. 2(e)).

\begin{figure}[tbp]
\centerline{\includegraphics*[width=7.2cm]{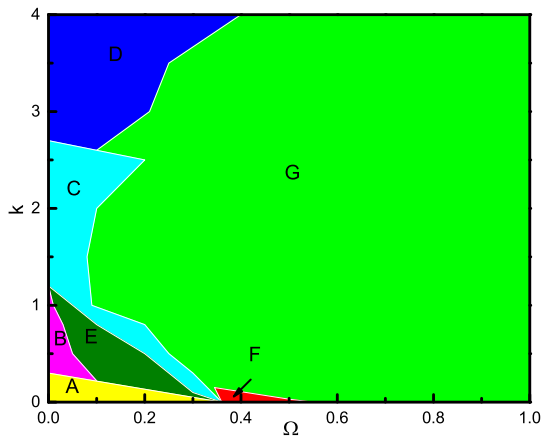}}
\caption{(Color online) Ground-state phase diagram of rotating spin-orbit
coupled spin-1/2 BEC in an optical lattice plus a harmonic trap with respect
to $\Omega $ and $k$ ($k_{x}=k_{y}=k$) for $\protect\beta =300$ and $\protect%
\beta _{12}=200$. There are seven different phases marked by A-G.}
\label{Figure3}
\end{figure}

Secondly, we give the ground-state phase diagram spanned by the rotation
frequency $\Omega $ and the isotropic SOC strength $k$ in Fig. 3. For the
case of rotating spin-orbit-coupled spin-$1/2$ BECs in a simple harmonic
trap only, previous investigations \cite{XQXu,Aftalion} showed that the
interplay between rotation frequency, SOC strength, and interparticle
interactions can lead to various ground-state phases, such as half-quantum
vortex, giant vortex, ringlike structures with triangular vortex lattices.
For the present system, there are seven different phases marked by A-G,
which differ in terms of their different density and phase distributions. In
the following discussion, we will give a description for each phase. We
start from the case where both the rotation and SOC are sufficiently weak,
which is indicated by the yellow region A in Fig. 3. This phase is the
periodically modulated TF phase without vortex in each component, which is
the same with the A Phase in Fig.1. The typical density and phase profiles
of the B phase, C phase and D phase in Fig. 3 are similar to those in Figs.
2(a), 2(b) and 2(c), respectively. At the same time, the density and phase
profiles of the phases E-G in Fig. 3 are shown in Fig. 7(a) and Figs.
4(a)-4(b), respectively.

\begin{figure}[tbp]
\centerline{\includegraphics*[width=8cm]{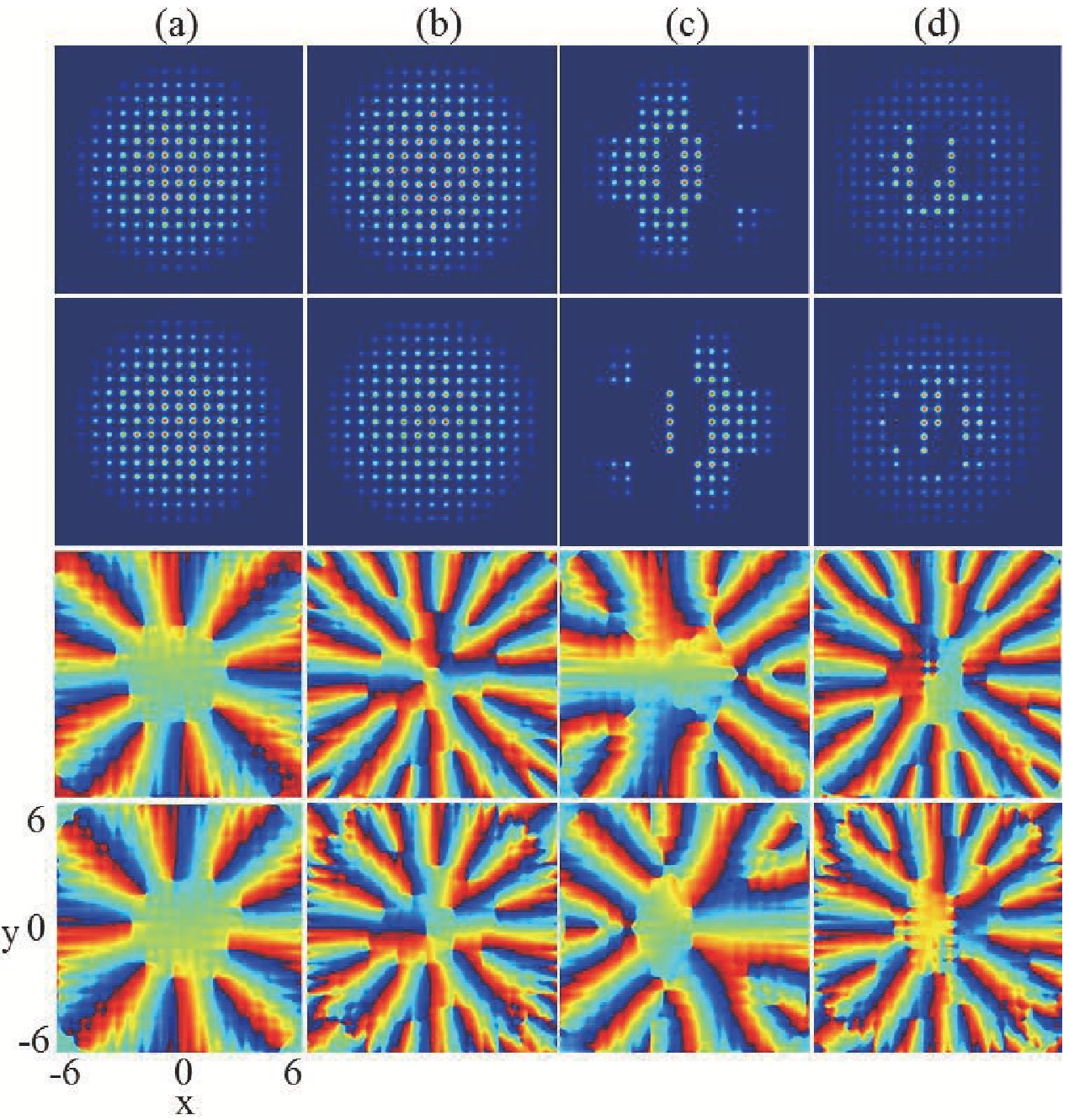}}
\caption{(Color online) Ground-state density distributions (the top two
rows) and phase distributions (the bottom two rows) for spin-1/2 BEC with
isotropic 2D RD-SOC and fixed rotation frequency $\Omega =0.5$ in an optical
lattice and a harmonic trap. (a) $k_{x}=k_{y}=0$, (b) $k_{x}=k_{y}=1$, (c) $%
k_{x}=k_{y}=0$, and (d) $k_{x}=k_{y}=1$. The interaction parameters are $%
\protect\beta =300$ and $\protect\beta _{12}=200$ for (a)-(b), and $\protect%
\beta =200$ and $\protect\beta _{12}=300$ for (c)-(d). The odd and even rows
correspond to component 1 and component 2, respectively. The unit length is $%
a_{0}$.}
\label{Figure4}
\end{figure}

For relatively large rotation frequency but very weak SOC, the system
exhibits a vortex ring where the vortices form a ring structure, indicated
by red region F in Fig. 3. The main results are illustrated in Fig. 4(a). In
Fig. 4, we consider the effect of isotropic 2D RD-SOC on the ground-state
structure of the system with fixed rotation frequency $\Omega =0.5$. For
initially miscible two-component BECs without SOC, i.e., $k_{x}=k_{y}=0$, a
vortex ring forms in each component and the density distributions are almost
the same (see Fig. 4(a)), i.e., the F phase emerges. When the SOC strength
further increases, the F phase transforms to the G phase, as shown in Fig.
3. The typical ground state is that more vortices occur in any of individual
components and these vortices tend to form a triangular vortex lattice,
where some of them enter the central region of the external potential (Fig.
4(b)). This G phase occupies the largest region of the ground-state phase
diagram in Fig. 3. By comparison, for the case of initially immiscible BECs
without SOC, the component densities are fully separated and the vortices
form an irregular vortex cluster, which is displayed in Fig. 4(c). With the
increase of SOC strength, e.g., $k_{x}=k_{y}=1$, the system exhibits partial
phase mixing in spite of the two components being separated initially, and
the vortices and the vortex-antivortex pairs in each component constitute
complex topological structure (see Fig. 4(d)), which is resulted from the
competition among the repulsive interatomic interaction, SOC, rotation and
the optical lattice.

Next, we move to the case of relatively small rotation frequency and weak
SOC. In this regime, the system sustains E phase, which is denoted by region
E in Fig. 3. Typical density and phase distributions of the E phase are
displayed in Fig. 7(a). Obviously, the ground state is the known
half-quantum vortex state, which is characterized by one vortex in one
component and no vortex in the other component (see the third and fourth
columns of Fig. 7(a)).

We find that the present system supports not only the line-like vortex
excitation with respect to the spatial degrees of freedom of the BECs but
also the point-like topological excitation (skyrmion excitation) concerning
the spin degrees of freedom. The skyrmion is a type of topological soliton,
which was originally suggested in nuclear physics by Skyrme to elucidate
baryons as a quasiparticle excitation with spin pointing in all directions
to wrap a sphere \cite{Skyrme}. It can be viewed as the reverse of the local
spin, which has been observed in many condensed-matter systems, such as
quantum Hall system, liquid crystals, helical ferromagnets, liquid $^{3}$%
He-A, and BECs \cite{Anderson2,Wright,XZYu,CFLiu}.\ The nonsingular skyrmion
in two-component BECs is related to the Mermion-Ho coreless vortices \cite%
{Mermin}, and\ the combination of SOC and rotation can cause various
topological defects including circular-hyperbolic skyrmion \cite{CFLiu},
giant skyrmion \cite{Aftalion}, and so on. In Fig. 5 we display the
topological charge density and the local enlargements of the spin texture
for the ground state in Fig. 4(a). Here the topological charge is expressed
by $Q=\int q(\mathbf{r})dxdy$ with the topological charge density $q(\mathbf{%
r})=\frac{1}{4\pi }\mathbf{S\bullet (}\frac{\partial \mathbf{S}}{\partial x}%
\times \frac{\partial \mathbf{S}}{\partial y}\mathbf{)}$\textbf{. }Our
numerical calculation shows that the local topological charges in Figs. 5(b)
and 5(c) approach $Q=0.5$, which indicates that the local topological
defects in Figs. 5(b) and 5(c) are circular half-skyrmion (meron) and
hyperbolic half-skyrmion, respectively. At the same time, the local
topological charges in Figs. 5(d)-5(h) approach $Q=1$, which means that the
local topological defects are circular-hyperbolic skyrmion [see Figs.
5(d)-5(f)], hyperbolic-radial(out) skyrmion [Fig. 5(g)] and
hyperbolic-radial(in) skyrmion [Fig. 5(h)], respectively. Therefore the spin
texture in Fig. 5(a) forms an exotic skyrmion-half-skyrmion lattice
(skyrmion-meron lattice) composed of circular-hyperbolic skyrmions,
hyperbolic-radial(out) skyrmion, circular half-skyrmions, and hyperbolic
half-skyrmions. Essentially, the interesting topological structure is
resulted from the interplay among the optical lattice, rotation and the
interatomic interactions.

\begin{figure}[tbp]
\centerline{\includegraphics*[width=15cm]{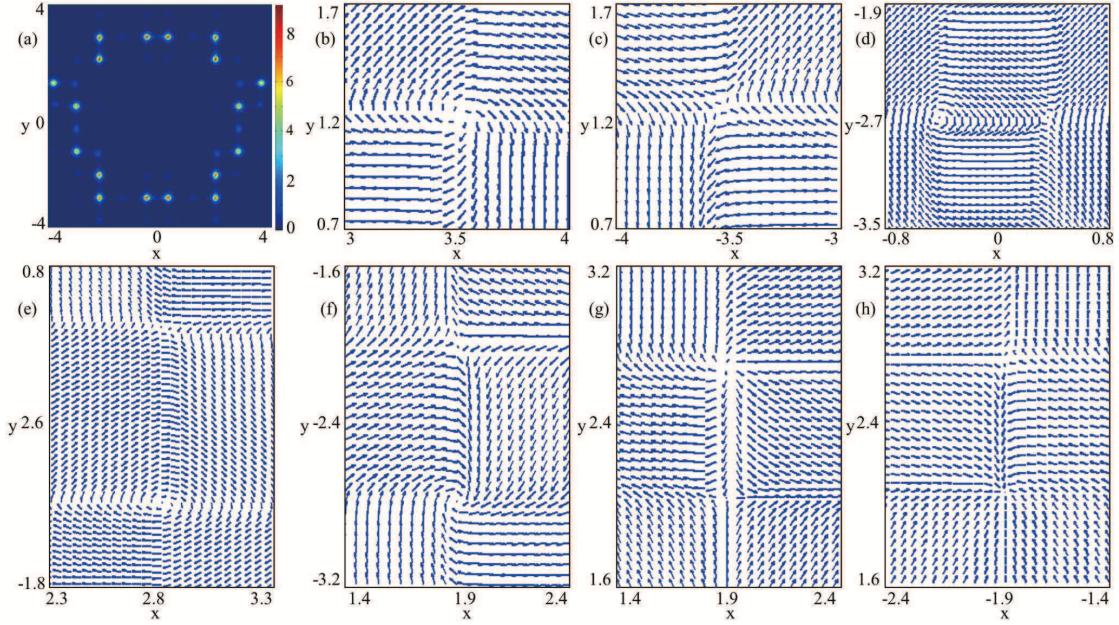}}
\caption{(Color online) Topological charge density (a) and local
enlargements of the spin texture (b-h) for rotating spin-1/2 BEC in an
optical lattice plus a harmonic potential, where $\protect\beta =300$, $%
\protect\beta _{12}=200$, $k_{x}=k_{y}=0$,$\ $and $\Omega =0.5$.The
corresponding ground state is shown in Fig. 4(a). The unit length is $a_{0}$%
. }
\label{Figure5}
\end{figure}

Figure 6(a) shows the topological charge density in the case of $\beta =300$%
, $\beta _{12}=200$, $k_{x}=k_{y}=1\ $and $\Omega =0.5$, where the
corresponding ground state is displayed in Fig. 4(b). The local spin texture
are given in Figs. 6(b)-6(d). Our computation results demonstrate that the
two local topological defects in Fig. 6(b) are a meron (half-skyrmion) pair
composed of two circular merons (half-skyrmions) with each local topological
charge being $Q=0.5$. In the mean time, the central meron pair is surrounded
by some other spin defects (the full spin texture is not shown here in view
of the limited resolution ratio and the brevity of the article). Our
simulation shows that the local topological charge for each of the outer
spin defects [e.g., see Figs. 6(c) and 6(d)] is $Q=0.5$, which implies that
these outer spin defects are merons (half-skyrmions). Thus the circular
meron pair and the half-skyrmions (merons) jointly form a complex meron
lattice (half-skyrmion lattice), which has not been reported in previous
literature. Physically, the asymmetry of the complicated topological
structure (i.e., the composite meron lattice) is caused by the destruction
of the SU(2) symmetry of the system in the presence of RD-SOC. The
interesting and exotic spin textures as mentioned in Figs. 5 and 6 allow to
be tested and observed in the future cold atom experiments.

\begin{figure}[tbp]
\centerline{\includegraphics*[width=8cm]{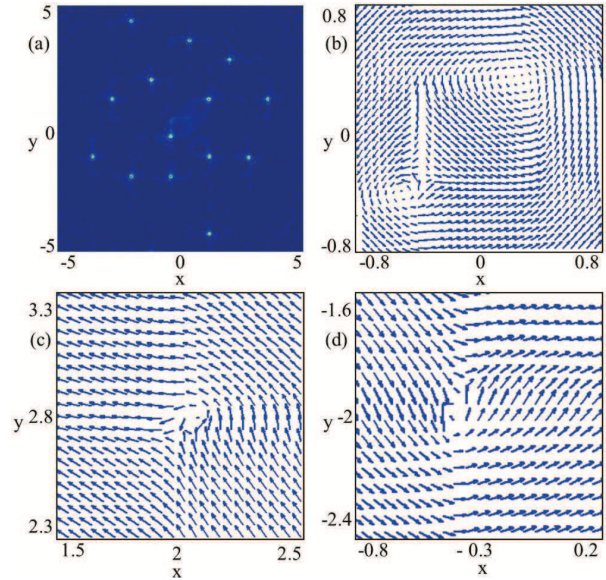}}
\caption{(Color online) Topological charge density (a) and local
enlargements of the spin texture (b-d) for rotating spin-1/2 BEC in an
optical lattice plus a harmonic potential, where $\protect\beta =300$, $%
\protect\beta _{12}=200$, $k_{x}=k_{y}=1$,$\ $and $\Omega =0.5$.The
corresponding ground state is given in Fig. 4(b). The unit length is $a_{0}$%
. }
\label{Figure6}
\end{figure}

Finally, we investigate the combined effects of RD-SOC, rotation and
interatomic interactions on the ground state of the system. Figure 7 shows
the density distributions and phase distributions for the ground states of
rotating two-component BECs with RD-SOC in an optical lattice and a harmonic
trap, where $k_{x}=k_{y}=0.5$. The rotation frequencies for the case of
initial phase mixing with $\beta =300$ and $\beta _{12}=200$ in Figs. 7(a)
and 7(b) are $\Omega =0.1$ and $\Omega =0.8$, and those for the case of
initial phase separation with $\beta =200$ and $\beta _{12}=300$ in Figs.
7(c) and 7(d) are $\Omega =0.3$ and $\Omega =0.8$, respectively. The columns
from left to right denote $\left\vert \psi _{1}\right\vert ^{2}$, $%
\left\vert \psi _{2}\right\vert ^{2}$, $\arg \psi _{1}$, $\arg \psi _{2}$,
and $\left\vert \psi _{1}\right\vert ^{2}-\left\vert \psi _{2}\right\vert
^{2}$, respectively.

\begin{figure}[tbp]
\centerline{\includegraphics*[width=12cm]{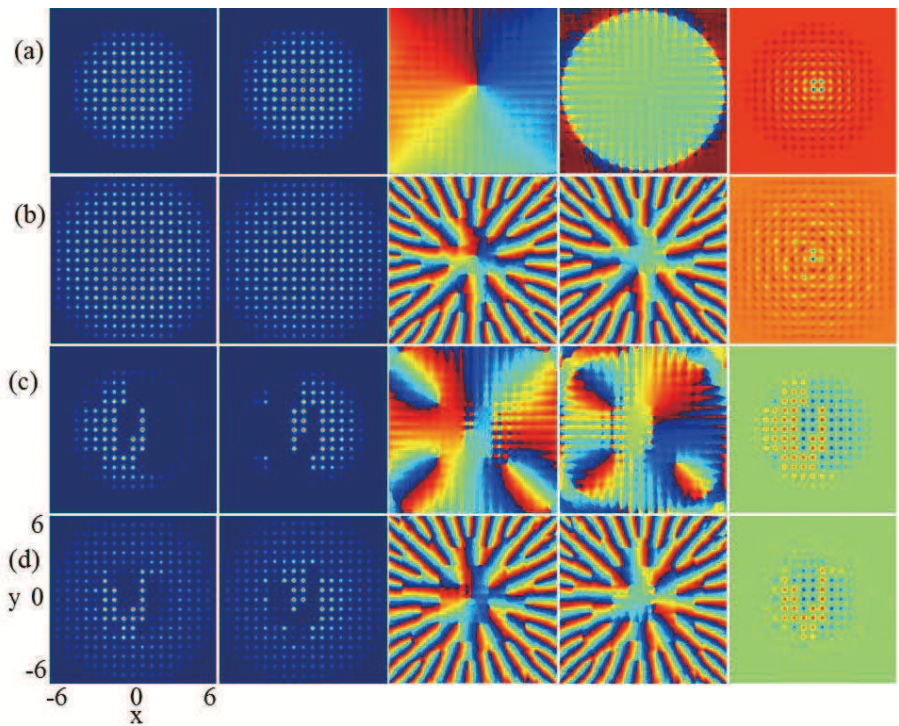}}
\caption{(Color online) Ground states of rotating 2D spin--1/2 BEC with
RD-SOC in an optical lattice and a harmonic trap, where $k_{x}=k_{y}=0.5.$
(a) $\Omega =0.1$, (b) $\Omega =0.8$, (c) $\Omega =0.3$, and (d) $\Omega
=0.8 $. The effective interaction parameters are $\protect\beta =300$ and $%
\protect\beta _{12}=200$ for (a)-(b), and $\protect\beta =200$ and $\protect%
\beta _{12}=300$ for (c)-(d). The columns from left to right represent $%
\left\vert \protect\psi _{1}\right\vert ^{2}$, $\left\vert \protect\psi %
_{2}\right\vert ^{2}$, $\arg \protect\psi _{1},$ $\arg \protect\psi _{2}$,
and $\left\vert \protect\psi _{1}\right\vert ^{2}-\left\vert \protect\psi %
_{2}\right\vert ^{2}$, respectively. The unit length is $a_{0}$.}
\label{Figure7}
\end{figure}

For the case of initial component mixing, as the rotation frequency
increases from zero, the ghost vortices on the outskirts of the atom cloud
enter the condensates and become visible vortices, where the phase defects
tend to form a triangular vortex lattice for large rotation frequency [see
Fig. 2(a), Fig. 7(a) and Fig. 7(b)] and the energy of the system reaches the
minimum in the rotating frame. For the case of initial component separation,
with the increase of rotation frequency, the topological structure of the
system gradually evolves from a square vortex lattice composed of
vortex-antivortex pairs into a complex triangular vortex lattice made of
pure vortices, and the two component densities transform from full phase
separation into partial phase mixing [see Fig. 2(d), Fig. 7(c) and Fig.
7(d)], which is quite different from the usual prediction results in
rotating two-component BECs with or without SOC \cite%
{XQXu,XZhou,HHu,Aftalion,Fetter,Kasamatsu}. In addition, we find that the
higher the rotation frequency is, the more the component densities of the
BECs expand. This point can be understood. Physically, when the rotation
frequency (with fixed RD-SOC and other parameters) increases, more angular
momentum contributes to the system and leads to the creation of more phase
defects and the expansion of the atom cloud, regardless of the initial state
of the system being mixed or separated.

\section{Ground-state structures and spin textures for the case of 1D RD-SOC}

Now, we consider the ground-state structures of rotating two-component BECs
loaded in an optical lattice and a harmonic trap in the presence of 1D
RD-SOC. From Fig. 8, we see that the larger the 1D SOC strength or the
rotation frequency is, the stronger the 1D SOC effect becomes. Taking the
case of fixed 1D RD-SOC strength with $k_{x}$ $=0$ and $k_{y}=2$ [see Figs.
8(c) and 8(d)] as an example, we first aim to discuss the influence of the
rotation frequency on the ground-state structure of the system. Figs. 8(c)
and 8(d) display the ground states of the system with $\Omega =0.3$ and $%
\Omega =0.8$, respectively. For the low rotation frequency $\Omega =0.3$,
there is an obvious visible vortex chain along $x=0$ axis in each component
due to the 1D RD-SOC along the $y$ direction [see columns 3 and 4 from left
to right in Fig. 8(c)], where the two component densities exhibit partial
mixing and partial separation. As the rotation frequency increases to $%
\Omega =0.8$, more vortices are generated along the $x=0$ axis and the both
sides of the central vortex chain, where the two component densities shows
good phase mixing except that the densities along the $x=0$ axis display
obvious phase separation [see columns 3 and 4 from left to right in Fig.
8(d)]. The physical mechanism is that large rotation frequency generates
more energy and more angular momentum. Thus the $x$-direction vortex chain
caused by the combined effect of 1D RD-SOC and rotation can only carry
finite energy and angular momentum, and the remaining energy and angular
momentum are inevitably carried by the transverse vortices beside the $x=0$
axis.

\begin{figure}[tbp]
\centerline{\includegraphics*[width=12cm]{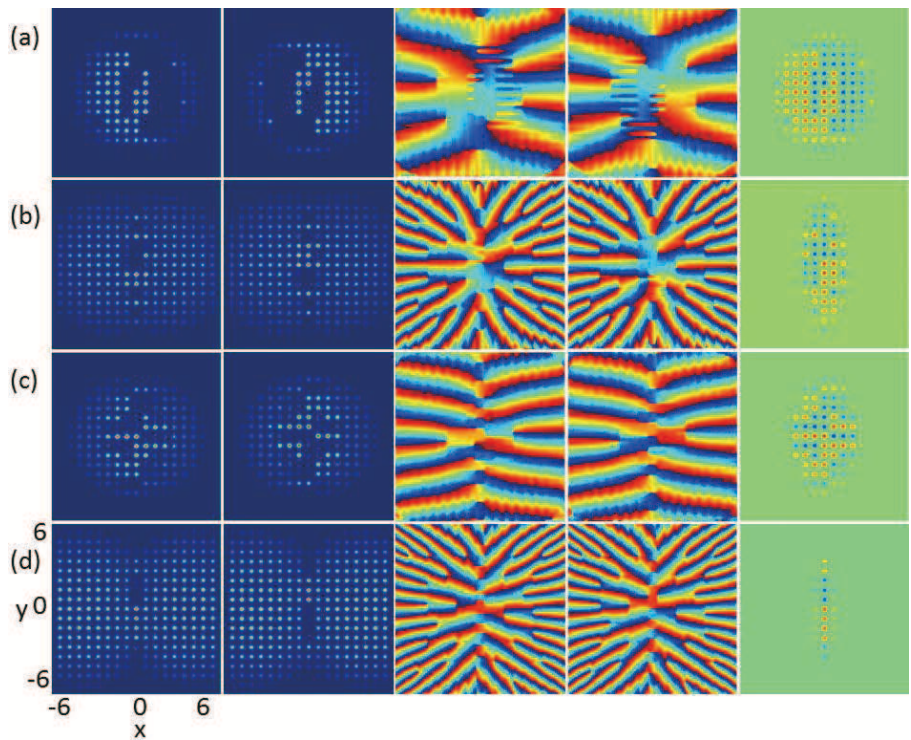}}
\caption{(Color online) Ground states of rotating spin-1/2 BEC with 1D
RD-SOC in an optical lattice and a harmonic trap, where $\protect\beta =200$%
, and $\protect\beta _{12}=300$. (a) $k_{x}=0$, $k_{y}=1$, $\Omega =0.3$,
(b) $k_{x}=0$, $k_{y}=1$, $\Omega =0.8$, (c) $k_{x}=0$, $k_{y}=2$, $\Omega
=0.3$, and (d) $k_{x}=0$, $k_{y}=2$, $\Omega =0.8$. The columns from left to
right represent $\left\vert \protect\psi _{1}\right\vert ^{2}$, $\left\vert
\protect\psi _{2}\right\vert ^{2}$, $\arg \protect\psi _{1},$ $\arg \protect%
\psi _{2}$, and $\left\vert \protect\psi _{1}\right\vert ^{2}-\left\vert
\protect\psi _{2}\right\vert ^{2}$, respectively. The unit length is $a_{0}$%
. }
\label{Figure8}
\end{figure}

Then we consider the influence of 1D RD-SOC on the ground-state structure of
the system. For instance, the rotation frequency is fixed as $\Omega =0.3$.
From Figs. 8(a) and 8(c), it is shown that the stronger 1D RD-SOC enhances
the creation of vortex chain and the formation of phase mixing. Similarly,
for the BECs with 1D RD-SOC along the $x$ direction ($k_{y}$ $=0$), our
simulation demonstrates that the density modulation occurs only along the $x$
direction. The above phenomena can be obtained and understood if one
performs an unitary transformation, $\sigma _{x}\rightarrow \sigma _{y}$ and
$\sigma _{y}\rightarrow -\sigma _{x}$, and sets $k_{x}$ or $k_{y}=0$ for the
RD-SOC term.

The topological defects can be observed in a phase profile, but a better
visualization is to use the pseudospin representation based on Eqs. (\ref%
{SpindensityX})-(\ref{SpindensityZ}). We can plot the functions $S_{x},$ $%
S_{y}$ and $S_{z}$ which reveal the presence of all the spin defects. The
corresponding spin-density distributions of Figs. 8(a)-8(d)\ are shown in
Figs. 9(a)-9(d), respectively. In the pseudo-spin representation, the red
region denotes $1$ (spin-up) and the blue region denotes $-1$ (spin-down).
According to Eq. (\ref{SpindensityZ}), the spin-density component $S_{z}$ is
related to the density difference of the two components, therefore the
variation tendency of the last row of Fig. 9 is consistent with\ that of the
last column of Fig. 8. $S_{x}$ and $S_{y}$ obey neither even parity
distribution nor odd parity distribution along the $x$ direction or the $y$
direction.

\begin{figure}[tbp]
\centerline{\includegraphics*[width=9cm]{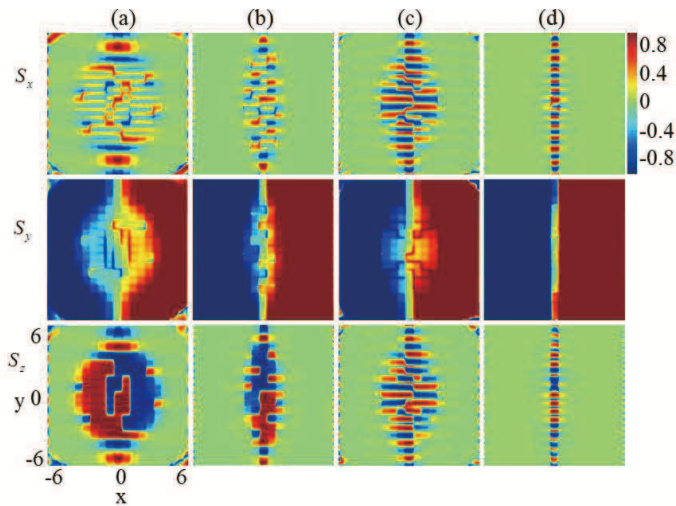}}
\caption{(Color online) Spin densities of rotating spin-1/2 BECs with 1D
RD-SOC in an optical lattice and a harmonic trap, where $\protect\beta =200$
and $\protect\beta _{12}=300$. (a) $k_{y}=1$,\ $\Omega =0.3$,\ (b) $k_{y}=1$%
,\ $\Omega =0.8$, (c) $k_{y}=2$,\ $\Omega =0.3$, and (d) $k_{y}=2$,$\ \Omega
=0.8$. The rows from top to bottom denote $S_{x},$ $S_{y}$ and $S_{z}$
components of the spin density vector, respectively. The corresponding
ground states for (a)-(d) are shown in Figs. 8(a)-8(d), respectively. The
unit length is $a_{0}$. }
\label{Figure9}
\end{figure}

For further comparison, we choose Fig. 9(a) ($\Omega =0.3\ $and $k_{y}$ $=1$%
)\ and Fig. 9(c) ($\Omega =0.3$ and $k_{y}$ $=2$) as an example, and we can
see that for the region of $x>0$ ($x<0$) the value of $S_{y}$ gets larger
(smaller) with the increasing $k_{y}$. Similarly, from Figs. 9(c) and 9(d),
we can observe that for the region of $x>0$ ($x<0$) $S_{y}$ approaches $1$ ($%
-1$) when the rotation frequency $\Omega $ increases (see the middle row of
Fig. 9).\ Thus we conclude that the spin component $S_{y}$\ develops into
two remarkable spin domains due to the increase of $k_{y}$ or $\Omega $. At
the same time, an obvious spin domain wall forms on the boundary region
between the two spin domains, which can be seen in the middle row of Fig. 9.
In general, the spin domain wall for nonrotating two-component BECs is a
classical N\'{e}el wall, where the spin flips only along the vertical
direction of the wall. Whereas, our numerical simulation of the spin texture
demonstrates that the spin in the region of spin domain wall flips not only
along the vertical direction of domain wall (the $x$ direction) but also
along the domain-wall direction (the $y$ direction), which indicates that
here the domain wall is an unique Bloch wall instead of the conventional N%
\'{e}el wall. This domain wall is the product of the phase-separated
two-component BECs in response to external rotation or RD-SOC, and reflects
the influence of rotation or RD-SOC on the magnetism of BECs.

\begin{figure}[tbp]
\centerline{\includegraphics*[width=12cm]{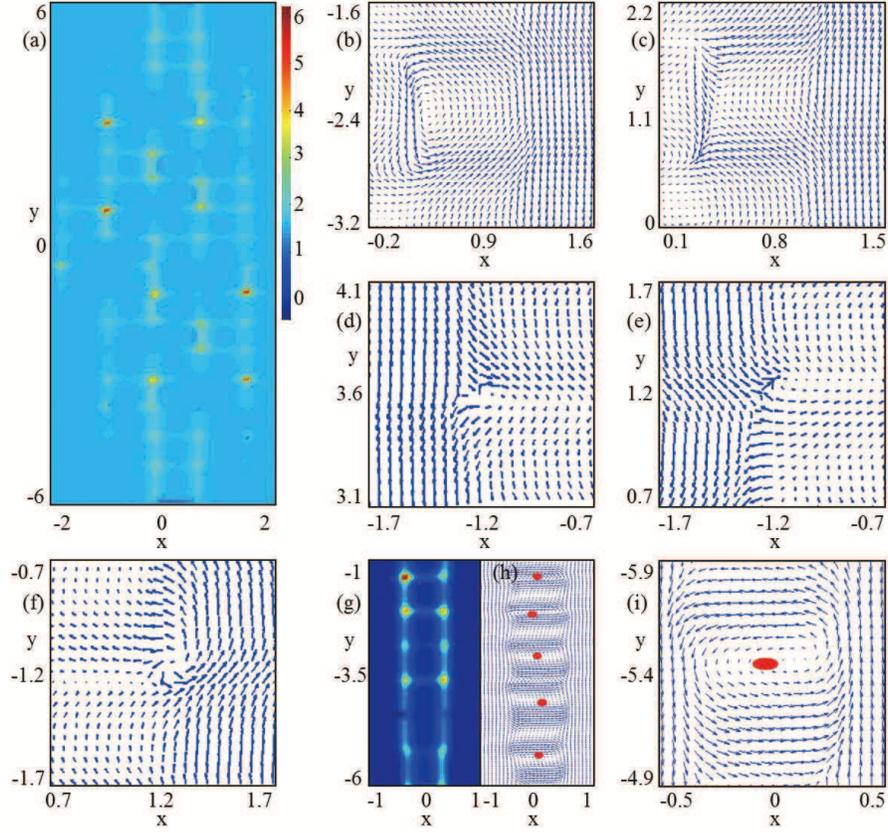}}
\caption{(Color online) Topological charge densities and spin textures of
rotating spin-1/2 BECs with RD-SOC in an optical lattice and a harmonic
trap. (a) topological charge density, and (b)-(c) local amplifications of
the spin texture, where the ground state is given in Fig. 8(b). (d)
topological charge density, (e) the corresponding spin texture, and (f)
local enlargement of the spin texture in (e), where the ground state is
given in Fig. 8(d). The unit length is $a_{0}$.}
\label{Figure10}
\end{figure}

Displayed in Fig. 10(a) is the topological charge density for the parameters
in Fig. 8(b). The typical local spin textures are shown in Figs.
10(b)-10(f). Our computation results show that the local topological charges
in Figs. 10(b) and 10(c) are both $Q=1$ while those in Figs. 10(d)-10(f) are
all $Q=0.5$. Thus the spin defects in Figs. 10(b) and 10(c) denote an
irregular circular skyrmion and an irregular hyperbolic skyrmion \cite%
{Kasamatsu,CFLiu}, respectively. At the same time, the topological defects
in Figs. 10(d)-10(f) represent circular half-skyrmion (meron) \cite{Mermin},
hyperbolic half-skyrmion and circular half-skyrmion, respectively. These
topological defects alternately appear in the spin representation of Fig.
8(b) and constitute a complicated skyrmion-half-skyrmion (skyrmion-meron)
lattice. In addition, we find that for strong 1D RD-SOC the system favors an
exotic skyrmion chain (i.e., the skyrmions form a chain-like structure)
which traverse the BECs. Figure 10(g) shows the topological charge density,
where the ground state is given in Fig. 8(d). Figure 10(h) gives the
corresponding spin texture, and the typical local amplification is exhibited
in Fig. 10(i). Our numerical calculation shows that the local topological
charge in Fig. 10(i) is $Q=1$ and the total topological charge in Fig. 10(h)
is $Q=5$. Therefore the spin structure of the system is a skyrmion chain
that is composed of a string of elliptic skyrmions with unit topological
charge in spin space as shown in Fig. 10(h). Obviously, the skyrmion
configurations observed in the present system are remarkably different from\
the previously reported results in rotating two-component BECs with or
without SOC \cite{XQXu,XZhou,HHu,Aftalion,Fetter,Kasamatsu,CFLiu}.

\section{Conclusion}

In summary, we have investigated the topological excitations of rotating
two-component BECs with RD-SOC in a 2D optical lattice and a 2D harmonic
trap. The effects of 2D RD-SOC, 1D RD-SOC, rotation, interatomic
interactions and optical lattice on the topological structures of the ground
states of the system are systematically discussed. Two ground-state phase
diagrams for the nonrotating case and the rotating case are given with
respect to the SOC strength and the interspecies interaction, and with
respect to the rotation frequency and the SOC strength, respectively.
Without rotation, a relatively weak isotropic 2D RD-SOC induces the
formation of rectangular vortex-antivortex lattice for initially separated
BECs, but strong 2D RD-SOC leads to generation of vortex-antivortex rings.
For fixed isotropic 2D RD-SOC strength, the depth of optical lattice and the
interaction parameters, the increase of rotation frequency can trigger the
structural phase transition from square vortex lattice to irregular
triangular vortex lattice and the system evolves from initial phase
separation into phase mixing. For the case of 1D RD-SOC, the increase of SOC
strength or rotation frequency may result in the creation of vortex chain
and Bloch domain wall. In addition, the system sustains novel spin texture
and skyrmion structures including an exotic skyrmion-half-skyrmion lattice
(skyrmion-meron lattice), a complicated meron lattice (half-skyrmion
lattice) and a skyrmion chain. These new topological excitations are quite
from the predictions in the previous literature of rotating two-component
BECs with or without SOC. Theoretically, the rotating spin-orbit-coupled BEC
in an optical lattice plus a harmonic trap is feasible and can be achieved
in principle. The experimental realization of the model Hamiltonian Eq. (\ref%
{Hamiltonian}) is still a challenge within the current experimental
conditions. However, considering that the present system has various novel
physical properties, with the continuous development of experimental
techniques, the system may be implemented in the future and its novel
topological excitations are expected to be observed in experiments.

\begin{acknowledgments}
L.W. thanks Professor Zhaoxin Liang and Professor Chaofei Liu for helpful
discussions, and acknowledges the research group of Professor W. Vincent Liu
at The University of Pittsburgh, where part of the work was carried out.
This work was supported by the National Natural Science Foundation of China
(Grant Nos. 11475144 and 11047033), the Natural Science Foundation of Hebei
Province of China (Grant Nos. A2019203049 and A2015203037), and Research
Foundation of Yanshan University (Grant No. B846).
\end{acknowledgments}

\end{document}